\documentclass[12pt]{iopart}

\usepackage{iopams}
\usepackage{dsfont}
\usepackage{graphicx}
\begin{document}

\title[]{Estimating the spectral density of unstable scars}

\author{D Lippolis}

\address{Institute for Applied Systems Analysis, Jiangsu University, Zhenjiang 212013, China}
\ead{domenico@ujs.edu.cn}
\vspace{10pt}
\begin{indented}
\item[]June 2022
\end{indented}

\begin{abstract}
In quantum chaos, spectral statistics generally follows the predictions of Random Matrix Theory (RMT).
A notable exception is given by scar states, that enhance probability density around unstable periodic orbits 
of the classical system, therefore causing significant deviations of the spectral density from RMT expectations.
In this work, the problem is considered of both RMT-ruled and scarred chaotic systems coupled to an opening.
In particular, predictions are derived for the spectral density of a chaotic Hamiltonian scattering into a single-
or multiple channels. The results are tested on paradigmatic quantum chaotic maps on a torus. The present report 
develops the intuitions previously sketched in [D. Lippolis, EPL \textbf{126} (2019) 10003].     

\end{abstract}

%
%
\submitto{\JPA}
%
%
%

\section{Introduction}

Quantum chaotic systems are generally characterized by apparent randomness of wavefunctions, spectra, and other observables, and yet, they sometimes 
feature remarkable signatures of their underlying classical Hamiltonians~\cite{Haake}. 
A notable example is given by scars~\cite{Hel84}, that is eigenstates  with enhanced or suppressed probability density with respect to the predictions of Random Matrix Theory (RMT)~\cite{Mehta},
which instead models Hamiltonians as random matrices constrained by some symmetries, having waves of random amplitudes and phases 
as eigenfunctions. When first observed in the Bunimovich stadium, scars came as a surprise because they belong with quantum stationary states, while they are in fact
concentrated around \textit{unstable} periodic orbits of the underlying classical system, and thus would have no reason to `survive' the quantization.
Scarred states have been thoroughly studied in a number of models~\cite{Nonne03,Bogo04,Bogo06,Carlo16,Revuelta20}, and they have finally been ascribed to a mechanism of constructive interference~\cite{Bogo88}, although very recent results suggest that scarring may be a classical effect, after all~\cite{Lipp21}.
Experimentally, they have mainly been observed in microwave billiards~\cite{stoeck}, as well as in optical microcavities~\cite{An02,Taka03,CaoWier}. When dealing with these platforms, scar states are 
generally identified by looking for closed orbits of the classical billiard that bears the same shape of the resonator at hand, which is, however, an \textit{open}
system, whose spectrum is made of resonances of finite widths, and unstable eigenfunctions with decaying rates and radiant patterns.
Hence the idea of studying scars in the framework of scattering. 
The dynamics and statistics of unstable quantum states in chaotic systems is one the leading themes of nuclear spectroscopy, where a number of 
predictions have been derived for such observables as the scattering matrix, all based on RMT~\cite{Weiden86,SokZel89,SokZel92,ISSO94,FyodSav12}.
A milestone in that context is certainly the effect of `collectivization of widths'~\cite{SokZel89,SokZel92}, that is the losses due to the leak/scattering process are approximately equally
distributed among the widths of all resonances, with the exception of a few superradiant states, equal in number to the open channels.
The phenomenon has been observed in several different models, and later redubbed `resonance trapping'~\cite{rotlet}.
Appealing as they are, the ensemble averages leading to the predictions mentioned above cannot na\"ively be performed to deal with chaotic scattering if
scar states are involved, due to the significant deviations of the spectral statistics from RMT, and thus the expectations previously obtained for the scattering 
matrix  and similar observables would break down.

In the present report, an attempt is made to fill this gap.           
The spectral density, a key observable that can be practically measured in physical experiments, is predicted for a scarred system
coupled to single- and multiple-channel openings. The theoretical expectations are obtained starting from the semiclassical evolution of wavepackets
in the neighborhood of the periodic orbits giving birth to scars, that allows us to locally estimate the autocorrelation function for the closed system.
Then, a self-consistent equation is used to work out an expression for the Green's function of the open system in terms of that of the closed system, and 
of the open channels. Tests follow on quantum maps with different time-reversal symmetries, coupled to single- and multichannel openings inspired by
microresonators.

The article is structured as follows: the local density of states is introduced in section~\ref{LDoS}, together with our model of choice for open systems
stemming from a chaotic Hamiltonian. Existing theories for the estimation of the local density of states in the realm of closed systems are exposed    
in detail in section~\ref{LdosClosed}. Both RMT expressions and semiclassical treatment of the problem in the presence of scars are  thoroughly reviewed.
The present approach to predict spectral densities in chaotic scattering is introduced in section~\ref{openldos}, while closed-form expressions for both the
quantities local density of states and reactance, respectively imaginary and real part of the spectral density, are derived in section~\ref{1cpreds} for
a single-channel opening.
It is observed that this rather simple model features interesting behavior of the spectral density when scars are involved, and the expression derived 
has a surprisingly elegant form that is then retrieved in the results of  multichannel theory. The single-channel theory is successfully tested against numerics
in section~\ref{numer}, using an open quantized cat map as model. The multiple-channel theory is treated in section~\ref{mchan}, together with the 
time-reversal symmetric quantum map opened `\`a la Fresnel', aimed at mimicking the dynamics in a dielectric microresonator.
In two separate instances of $i)$ the opening away from the scar and thus coupling mainly with the RMT-governed states (section~\ref{scaraway}) and
$ii)$  the opening overlapping with the scar and thus mainly coupled to the scarred states (section~\ref{inscar}), a mean-field theory is developed 
that leads to closed-form expressions for the spectral density, formally similar to those obtained in the single-channel theory, all of them 
tested against the numerics. Conclusions and prospects follow suit.

\section{Local density of states}
\label{LDoS}
In absence of degeneracy, the local density of states for a Hamiltonian $H_0$ (closed system) is the Fourier transform of the correlation amplitude of an arbitrary quantum state $|x\rangle$  
\begin{equation}
S_0(E,x)  = \mathrm{Re}\,\int dt\langle x|U_0^t|x\rangle e^{iEt/\hbar}   
 = \sum_n |\langle x|\psi_n\rangle|^2\delta(E-E_n),
\label{ldos}
\end{equation}       
where $U_0^t=e^{-iH_0t/\hbar}$ is the quantum propagator of $H_0$.

On the other hand,
a typical scattering problem can be identified  with an effective Hamiltonian of the form~\cite{Fesh}
\begin{equation}
H = H_0  + \sum_c\int dE' \frac{H_0|\xi_c'\rangle\langle\xi_c'|H_0}{E-E'},  
\label{Heff}
\end{equation}
where the exit from the system may occur through a series of channels $|\xi_c\rangle$.
For chaotic Hamiltonians, the dependence of $H_0|\xi_c'\rangle$ on $E'$ is rather smooth and 
can be neglected if the energy range in exam is small enough~\cite{ISSO94}.
In that case, the principal part of the above integral can be reabsorbed in the Hermitian part of the model, $H_0$, while
the residue gives birth to a non-Hermitian term. The whole thing can be written as
\begin{equation}
H = H_0 -i\pi \sum_c H_0|\xi_c\rangle\langle\xi_c|H_0
\,.
\label{HnonH}
\end{equation}
The eigenenergies of this Hamiltonian have the form $\varepsilon_n-i\gamma_n$, where the imaginary part
represents the linewidth, or the lifetime of mode $n$. Moreover and importantly, $H$ has distinct left  
($\langle\Phi_n|$) and right ($|\Psi_n\rangle$) eigenfunctions. A generalization of the local density of states to this problem may thus be written as
\begin{eqnarray}
\fl
\nonumber
S(E,x) =\frac{1}{\hbar} \mathrm{Re} \int_0^\infty dt \langle x|U^t|x\rangle e^{iEt/\hbar} 
= \frac{1}{\hbar}\mathrm{Re} \frac{\langle x|\Psi_n\rangle\langle\Phi_n|x\rangle}{\langle\Phi_n|\Psi_n\rangle}
\int_0^\infty dt  e^{-i(\varepsilon_n-E)t/\hbar-\gamma_nt/\hbar}
\\ 
=  \mathrm{Re} \sum_n h_n(x)\frac{\gamma_n+i(E-\varepsilon_n)}{\gamma_n^2+(E-\varepsilon_n)^2}.
 \label{opnldos}
\end{eqnarray}
 Here $h_n(x)=  \frac{\langle x|\Psi_n\rangle\langle\Phi_n|x\rangle}{\langle\Phi_n|\Psi_n\rangle}$
 is the non-Hermitian counterpart of the intensity $|\langle x|\psi_n\rangle|^2$ appearing in Eq.~(\ref{ldos}) for the Hamiltonian of a closed system. The importance of this factor when characterizing localization patterns was already noted in Ref.~\cite{ErCaSa09}, in the context of quantum chaos.

 \subsection{Local density of states and chaos in closed systems}
 \label{LdosClosed}
 We briefly summarize well-known expectations for the local density of states in a chaotic system.
 First, consider a Hamiltonian whose spectral statistics follows Random Matrix Theory (RMT)~\cite{Haake}. If there are no deviations from RMT, the eigenfunctions may also be written as a superposition of plane waves with random amplitudes and phases, completely uncorrelated from the energy spectrum~\cite{BerTab77}.   
\begin{figure}
\centerline{
(a)
\includegraphics[width=5cm]{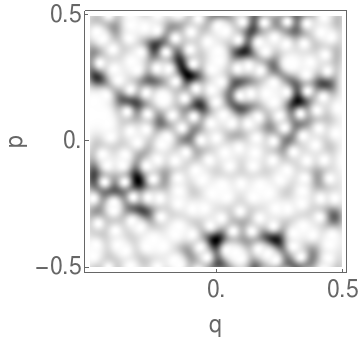}
(b)
\includegraphics[width=5cm]{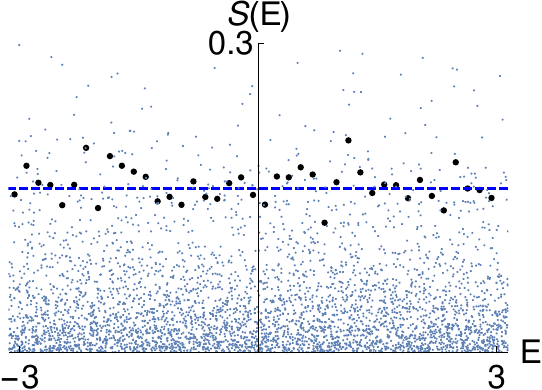}
(c)
\includegraphics[width=5cm]{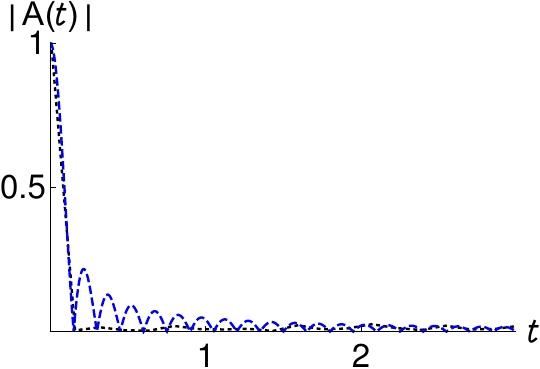}
}
\caption{Closed system: (a) The Husimi distribution of a generic (i.e. non-localized) eigenstate of the quantized cat map defined in~(\ref{eq:veps}). (b) The local density of states of the same cat map. Tiny dots: intensities $|\langle x|\psi\rangle|^2$ from 50 realizations of the quantum propagator with dimension from 200 to 300 (even numbers), and probe state chosen at random, but away from short periodic orbits. Dots: the normalized distribution of the intensities. Dashed line: the uniform distribution expected from COE statistics. (c) The autocorrelation function evaluated numerically (dotted line) as Fourier transform of the dots in (b) and the theoretical prediction [Eq.~(\ref{COEA}), dashed line] from the envelope of the local density of states.
}
\label{closednoscar}
\end{figure} 
Therefore, if we were to predict the overall envelope of the local density of states~(\ref{ldos}), we may just separate the spectral average from the average over intensities, and use well-known results from RMT to obtain~\cite{Mehta}
\begin{equation}
\fl
\overline{S}_0(E,x)  = \overline{|\langle x|\psi_n\rangle|^2}\,\overline{\sum_n\delta(E-E_n)} =
\left\{ \begin{array}{r} \frac{2}{\pi b^2}\sqrt{b^2-E^2}\theta(b^2-E^2) \hspace{1cm} \mathrm{GOE}
 \\ \frac{1}{2\pi} \hspace{2cm} \mathrm{COE} \end{array} \right.
 \,,
\label{avldos}  
 \end{equation}
assuming that the energy levels are located within the $(-b,b)$ interval, in the case of the Gaussian Orthogonal Ensemble (GOE).   
While, as said, this prediction is known, and it is not particularly exciting, we want to take a step back and recall that the autocorrelation function is the Fourier-Laplace transform of the local density of states, so that, for example, the Circular Orthogonal Ensemble (COE) yields
\begin{equation}
\overline{ A}(t)  = \int_{-\pi}^{\pi}e^{-iEt}S_0(E)dE = \frac{\sin\pi t}{\pi t} 
\, .
\label{COEA}
\end{equation}
The point is that the autocorrelation function decays rapidly, once rescaled by $\hbar$. In order to concisely sketch the properties of chaotic wavefunctions and to compare those to dynamical observables  such as the local density of states and the autocorrelation function, let us introduce the time-honored 
Q-representation (or Husimi), to visualize the wave function in the phase space:
\begin{equation}
{\cal{H}}(q_0,p_0) = \langle x|\hat{\rho}|x\rangle = \left|\langle\psi|x\rangle\right|^2
\, ,
\label{Husimi}
\end{equation}    
where
\begin{equation}
\langle q|x\rangle = \left(\frac{1}{\pi\hbar^2}\right)^{1/4}e^{-(q-q_0)^2/2\hbar+ip_0(q-q_0)/\hbar}
\label{wpack}
\end{equation}
is a wave packet centered at $(q_0,p_0)$, while
 the last identity in Eq.~(\ref{Husimi}) is valid for a pure state $\left(\hat{\rho}=|\psi\rangle\langle\psi|\right)$.
That way, the Husimi distribution of a typical chaotic eigenfunction of a quantum map (details of the quantization are provided in Section~\ref{numer}) is shown in Fig.~\ref{closednoscar}(a): albeit the wavefunction intensity fluctuates in the phase space, there are no scars, and the corresponding local density of states [Fig.~\ref{closednoscar}(b)] exhibits no pattern, so that its average over the ensemble would yield a uniform distribution (this is indeed the case for quantum maps on a torus, whose statistics follow Dyson's ensembles~\cite{Haake}). Consequently, the autocorrelation function decays rapidly with time, according to a Sinc-like behavior [Fig.~\ref{closednoscar}(c)].       

The situation is quite different for a chaotic system where scars are present. 
\begin{figure}[tbh!]
\centerline{
(a)
\includegraphics[width=4.5cm]{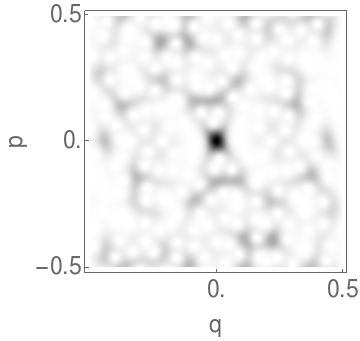}
(b)
\includegraphics[width=5cm]{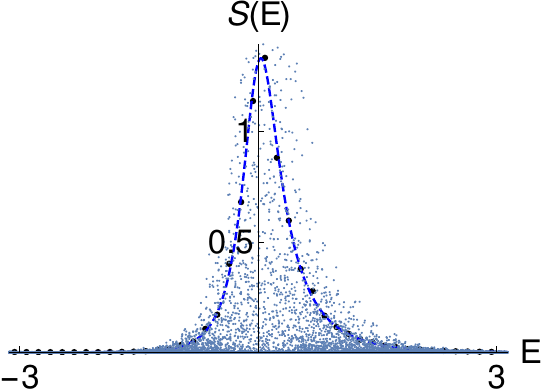}
(c)
\includegraphics[width=5cm]{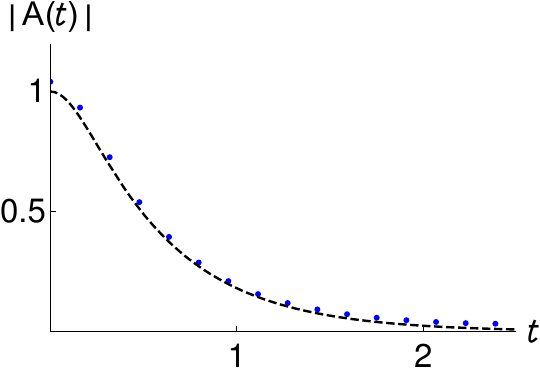}
}
\caption{Closed system: (a) The Husimi distribution of a scarred eigenstate of the quantized cat map. (b) Tiny dots: the intensity distribution of the same cat map vs. quasienenergy, computed numerically from 50 realizations of the quantum propagator with dimension from 200 to 300 (even numbers), and probe state centered at the origin, on the top of the fixed point. Dots: the normalized histogram of the intensities. Dashed line: the semiclassical prediction for the local density of states, defined by the Fourier transform of Eq.~(\ref{CatAp}). (c) The autocorrelation function evaluated numerically (dots) as Fourier transform of the histogram in (b), and (dashed line) the theoretical prediction~(\ref{CatAp}).}
\label{closedscar}
\end{figure} 
Enhancement of the probability density is evident as from the Husimi distribution of Fig.~\ref{closedscar}(a), in this case at the origin, in correspondence of the fixed point of the classical map. In addition to this feature, the local density of states [Fig.~\ref{closedscar}(b)] shows a non-trivial pattern, which largely deviates from RMT predictions, and, for a fixed-point scars, appears to have a peaked profile. 
Analogously, the autocorrelation function is also peaked [Fig.~\ref{closedscar}(c)], and it decays more slowly than its RMT-predicted counterpart from Fig.~\ref{closednoscar}(c).
It should be noted that here the test state $|x\rangle$ is the minimum-uncertainty wave packet~(\ref{wpack}) centered at the scar, but it is not an eigenstate of the Hamiltonian. 
Therefore, its \textit{long-time} evolution in a chaotic system is well approximated with a superposition of random waves extended over the whole space, and the autocorrelation function $\langle x|U_0^t|x\rangle$ asymptotically vanishes (its long-time limit is not exactly zero due to the non-orthogonality of wave packet and random wave). For that reason it has been argued in the past~\cite{Sred94} that the thermalization of generic initial states
is not affected by scarring.  
        
A semiclassical approximation to the envelope of both autocorrelation function and local density of states in the presence of scarring has been available for many years~\cite{Kap98,KapHel98}, based on the assumption that its main contribution comes from the propagation of the wavepacket on the manifolds, where its variance increases with the stability exponent of the periodic orbit. 
In the simplest case of orthogonal stable/unstable manifolds, the dynamics near the fixed point (here taken at the origin for simplicity) is linearized, and modelled as the Hamiltonian  
\begin{equation}
H = \lambda pq
\,.
\label{DL:fixp_haml}
\end{equation}
Quantizing this system amounts to write Schr\"{o}dinger's equation
\begin{equation}
i\hbar\partial_t \psi(q,t) = H\psi(q,t)
\,.
\label{DL:Sch1}
\end{equation} 
Assuming the unstable manifold coincides with the $q-$direction and the stable manifold with the $p-$direction, the previous equation may be written in coordinate space as 
\begin{equation}
\partial_t \psi(q,t) =  -\lambda\partial_q \left(q\psi(q,t)\right)
\,,
\label{DL:Sch2}
\end{equation}
that is Liouville equation for a density in the neighborhood of an expanding fixed point.
It's a particular situation in which the $i\hbar$'s in the equation cancel out,
but it turns out the result is similar to the semiclassical evolution of a wave packet in the
neighborhood of a hyperbolic unstable periodic orbit of finite period.
The corresponding integral evolution
operator is the Perron-Frobenius.
Equation~(\ref{DL:Sch2}) is solved by
\begin{equation}
\psi(q,t) = e^{-\lambda t} \psi_0\left(q e^{-\lambda t}\right)
\,,
\label{DL:solutn}
\end{equation}
so that the wave packet evolves as
\begin{equation}
e^{-q^2/2\sigma^2} \rightarrow e^{-\lambda t} e^{-q^2/2(\sigma e^{\lambda t})^2}
\,,
\label{DL:x_evol}
\end{equation}
meaning the width $\sigma$ of the packet stretches by a factor of $e^{\lambda t}$ after
a time $t$. Analogously, it can be shown that the width of the wavepacket contracts by a factor of $e^{-\lambda t}$ in momentum space. 
For a chaotic Hamiltonian approximated with~(\ref{DL:fixp_haml}) in the neighborhood of a fixed point $x_p$ of stability $\lambda_p$, 
the autocorrelation function
$A_p$ of a wavepacket $|x_p\rangle$ of the form~(\ref{DL:x_evol}) is estimated as 
\begin{equation}
A_p(t) = \langle x_p(0)|x_p(t)\rangle \simeq \sqrt{\frac{2\pi}{\cosh \lambda_p t}}
e^{iE_pt} 
\, .
\label{ScarA}
\end{equation}   
The semiclassical spectral density is just the Fourier transform of Eq.~(\ref{ScarA}):
\begin{equation}
G_{\mathrm{sc}}(E,x_p) = \int dt\,A_p(t)e^{iEt}
\,.
\label{Gsc}
\end{equation} 
The previous estimate only accounts for the so-called linear envelope, that is the one generated by short-time recurrences. As one can perhaps guess from Fig.~\ref{closedscar}, the local density of states evaluated numerically from the spectrum of the quantum map has a much finer structure. Particularly, one would still be able to semiclassically predict the ripples in the tails of the distribution with a `non-linear' theory that account for the homoclinic excursions from the periodic point, which eventually return to it.  A random hypothesis on the nonlinear part of the spectrum was formulated in~\cite{KapHel98}, whereas a closed-form semiclassical expression has been derived more recently~\cite{Verg12,Verg13}.
While one should be aware of such progress, we shall only use the linear envelope~(\ref{ScarA}) in what follows, as the starting point for the analysis of the dissipative system, while 
nonlinear effects will be accounted for on average by introducing a factor of $N^{-1}$ and thus approximating $A(t)\simeq A_p(t)/N$. As explained in detail in reference~\cite{KapHel98},
$N^{-1}$ is the average probability of recurrences in a chaotic system where wavefunctions are assumed random, and in our case it estimates the probability to travel between
any two regions of the phase space away from the neighborhood of the fixed point where the linear theory of scars  applies.       

\section{Local density of states in open systems}
\label{openldos}
Is it possible to estimate the local density of states of the open system form the statistical properties or a semiclassical approximation of the
spectrum of the closed system, plus the couplings with the continuum?

In order to find out, let us first
write the local density of 
states~(\ref{opnldos}) as
\begin{equation}
S(E,x) = \frac{1}{\hbar}\, \mathrm{Re} \int_0^\infty dt \langle x|U^t|x\rangle e^{iEt/\hbar}  
\, .
  \label{rewldos}
  \end{equation} 
On the other hand, the standard definition~\cite{Sakurai} of the Green's function is the Laplace transform of the expectation value of the quantum propagator [cf. Eq.~(\ref{opnldos})]:
\begin{equation}
\langle x|G(E)|x\rangle = -\frac{i}{\hbar}\int_0^\infty dt \langle x|U^t|x\rangle e^{iEt/\hbar} =
 \sum_n \frac{ h_n(x)}{E-\varepsilon_n + i\gamma_n}
 \, ,
\label{Gfunc}
\end{equation}
so that we have 
\begin{equation}
S(E,x) =  \mathrm{Im} \langle x|G(E)|x\rangle
\label{SG}
\end{equation}
In what follows, the Green's function will be used for the evaluation of $S(E,x)$.
Let us go back to the Hamiltonian~(\ref{Heff}), which may be simplified to the model~\cite{FyodSav12,Kap99}
\begin{equation}
H = H_0 -i\Gamma\sum_c|a_c\rangle\langle a_c| ,
\label{multcH}
\end{equation}
where $\Gamma$ is a dimensionless parameter that controls the coupling with the continuum. 
Call $W=\Gamma\sum_c|a_c\rangle\langle a_c|$.  
We may now use a self-consistent equation to express the Green's function in terms of the resolvent
of the closed system:
\begin{equation}
\fl
G(E) = G_0(E) - i\,G(E)\sum_c|a_c\rangle\langle a_c|G_0(E)  = 
 G_0(E) - i\,G(E)WG_0(E)
 \label{selfconst}
 \end{equation}
Then the Green's function can be formally expanded in $G_0(E)$ according to the 
above recursion relation,
to obtain~\cite{SokZel89,SokZel92}
\begin{eqnarray}
\fl
 \nonumber
G(E) = G_0(E) -  i\,G_0(E)WG_0(E)
-G_0(E)WG_0(E)WG_0(E) + ...
\\ 
\nonumber
= G_0(E) - i\,G_0(E)W\left[ 1  - i\,G_0(E)W - 
 G_0(E)WG_0(E)W +... \right]G_0(E)
\\ 
= G_0(E) -  i\,G_0(E)W \left[1 + i\,G_0(E)W\right]^{-1} G_0(E) .
\label{Dyson}
\end{eqnarray}
The previous is an expression that depends on the Green's function of the closed system and the open channels.

\subsection{Single-channel opening}
\label{1cpreds}
Although our initial goal is that to write the local density of states (or, equivalently, the Green's function)
of the open system in terms of its closed analog and the open channels, Eq.~(\ref{Dyson}) is not very insightful per se, and we would perhaps need a little more intuition to understand what dissipation does to scars. To that aim, the analysis is now restricted to a single-channel opening: the non-Hermitian Hamiltonian~(\ref{multcH}) becomes  
\begin{equation}
H = H_0 -i\Gamma |a\rangle\langle a| ,
\label{singcH}
\end{equation}
and the previous derivation in this case yields the Green's function
\begin{equation}
\langle x|G(E)|x\rangle = \langle x|G_0(E)|x\rangle -
 i\Gamma\frac{\left(\langle x|G_0(E)|a\rangle\right)^2}{1+i\Gamma\langle a|G_0(E)|a\rangle} 
 \, .
\label{Dysldos1c}
 \end{equation}
As in the multiple-channel result, the dependence of Eq.~(\ref{Dysldos1c}) on the spectral density of the probe state $|x\rangle$ in the closed system (the first term) and on  that of the open channel $\left(\langle a|G_0(E)|a\rangle\right)$ is apparent. In addition to that, though, the single-channel form reveals the `interaction' term $\left(\langle x|G_0(E)|a\rangle\right)^2$, which connects the probe state $|x\rangle$ to the open channel $|a\rangle$ through the quantum propagator. In order for the spectral density to be affected at all by the opening, that amplitude must be non-negligible. 

Let us consider the special case where $|a\rangle=|x\rangle$, meaning that the probe state is placed right on the top of the opening. Equation~(\ref{Dysldos1c}) further simplifies to
\begin{equation}
\langle x|G(E)|x\rangle = \frac{\langle x|G_0(E)|x\rangle}{1+i\Gamma\langle x|G_0(E)|x\rangle}
\label{ubersimpS}
\end{equation}
It is apparent from the previous expression that here one needs an estimate of the full Green's function of the closed system, which may be obtained in different ways, depending on whether the spectral statistics is random, or it exhibits localization. 
For the purpose of illustration we examine the two scenarios previously outlined in Section~\ref{LdosClosed}.
\begin{figure}[tbh!]
\centerline{
(a)
\includegraphics[width=7cm]{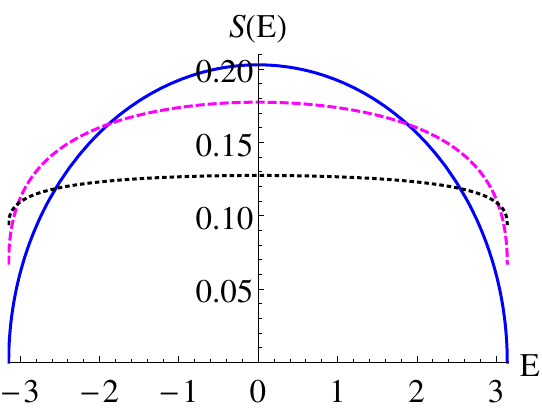}
(b)
\includegraphics[width=7cm]{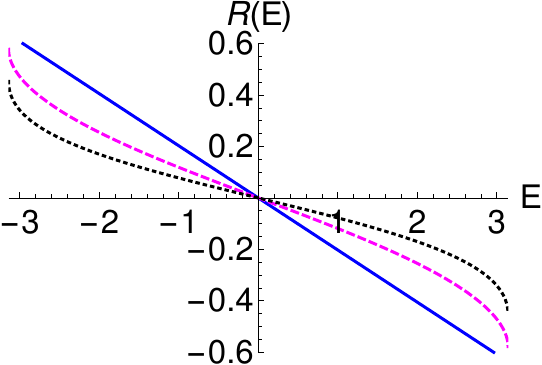}
}      
\caption{Prediction~(\ref{avG}) for the local density of states, with $|x\rangle=|a\rangle$ in the cases of: (a) a chaotic, non-scarred system to which GOE applies. Solid line: closed system (semicircle rule). Dashed line: open system, $\Gamma=0.5$. Dotted line: open system, $\Gamma=1$; (b) the real part of the Green's function, or reactance, for the same system.}
\label{onecpred}
\end{figure}

 First,  a chaotic system whose spectral and wavefunction statistics follow RMT is coupled to a single-channel opening. A prediction for the local density of states is given by Eq.~(\ref{ubersimpS}). 
 In order to estimate the spectral density of the closed system, $\langle x|G_0(E)|x\rangle$ in Eq.~(\ref{ubersimpS}),  
 it is appropriate in this case to take an average over the GOE akin to Eq.~(\ref{avldos}) to obtain
 \begin{equation}
 \fl
 g_{\mathrm{GOE}}(E) =
\overline{\langle x|G_0(E)|x\rangle}=\overline{\sum_n\frac{1}{(E-E_n)}\,|\langle x|\psi_n\rangle|^2} 
=\frac{2}{N\,b^2}\left[E-\sqrt{E^2-b^2}\right] 
\,,
\label{RMTav}
\end{equation}
here $(-b,b)$ is the range of energies in exam, $N$ the dimension of the Hilbert space. Analogously, one can show
(Appendix B) a similar result for the Circular Orthogonal Ensemble (COE), where quasienergies are angles $\theta$ in the range $(-\pi,\pi)$,
and the averaged spectral density must satisfy periodicity conditions:
\begin{equation}
g_{\mathrm{COE}}(\theta) =
\overline{\langle x|\frac{1}{1- \e^{i\theta}U}|x\rangle}= \frac{\e^{-i\theta}}{2N}
\,.
\label{gCOE}
\end{equation}
 At this point, one can use a mean-field approximation to  rewrite the averaged Eq.~(\ref{ubersimpS}) as
 \begin{equation}
\overline{G}(E,x) = \frac{\overline{\langle x|G_0(E)|x\rangle}}{1+i\Gamma\overline{\langle x|G_0(E)|x\rangle}}
\, .
\label{avG}
\end{equation}
The effect of the leak on the local density of states is  exemplified in 
fig.~\ref{onecpred}(a) for a system that follows GOE statistics.  
The outcome is that
the properly normalized local density of states $\frac{S(E)}{\int_{\Delta E}S(E)dE}$ tends to flatten and becomes increasingly close to a uniform distribution with the coupling to the leak. 
However, the semicircle rule [imaginary part of Eq.~(\ref{RMTav})] for the GOE and the uniform distribution for the COE, that describe the local density of states of  the closed system, 
have no peaks or notable structure in the first place. Therefore, the effect of the opening is that to make an already flat spectral density even flatter, and thus
qualitatively irrelevant for a system ruled by RMT.  

Secondly, we examine the more interesting case of a system that exhibits scarring, and couples the scar to an opening. As before, the prediction for the spectral density is given by Eq.~(\ref{ubersimpS}),
but this time $G_0(E)$ is expected to deviate from the RMT ensemble statistics, and to be energy-dependent, instead. Thus, rather than taking averages, one may well construct a semiclassical ansatz for the spectral density of the scarred system, on the basis of the available results already discussed in section~\ref{LdosClosed}.
Assuming a discrete, non-degenerate spectrum, 
\begin{eqnarray}
\fl
\nonumber
\langle x|G_0(E)|x\rangle &=& \sum_n|\langle x|\psi_n\rangle|^2 P\frac{1}{E-E_n} 
-i\pi\sum_n|\langle x|\psi_n\rangle|^2 \delta(E-E_n) \\ \nonumber
&=&  \sum_n|\langle x|\psi_n\rangle|^2\frac{i}{2}\int_{-\infty}^{\infty} du\,\mathrm{sgn}(u)\rme^{-i(E-E_n)u}
- \sum_n|\langle x|\psi_n\rangle|^2\frac{i}{2}\int_{-\infty}^{\infty}du\,\rme^{-i(E-E_n)u} \\ \nonumber
&=& -\frac{i}{2}\int_{-\infty}^{\infty}dv \sum_n|\langle x|\psi_n\rangle|^2\rme^{-iE_nv}\mathrm{sgn}(v)e^{iEv}
-  \frac{i}{2}\int_{-\infty}^{\infty}dv \sum_n|\langle x|\psi_n\rangle|^2\rme^{-iE_nv}\rme^{iEv} \\ 
&=&  -i\pi\int_{-\infty}^{\infty}dvA(v)\mathrm{sgn}(v)\rme^{iEv} 
- i\pi\int_{-\infty}^{\infty}dvA(v)\rme^{iEv}   \equiv  -R_0(E) + iS_0(E)
\, .
\label{reacldos}
\end{eqnarray}
That way the expectation value of the Green's function is written in terms of the autocorrelation function $A(t)$, that
can be approximated with $A_p(t)/N$, where $A_p(t)$ is the semiclassical ansatz for the autocorrelation function near the periodic orbit $p$, as for example in Eq.~(\ref{ScarA}), while the
factor of $N^{-1}$ is introduced to account for the `nonlinear' effects~\cite{KapHel98}. 
Since $A(-t)=A^*(t)$ [and $A_p(-t)=A_p^*(t)$], the two terms of Eq.~(\ref{reacldos})
are respectively real and pure imaginary. While the latter is identified with the local density of states, the former [$R_0(E)$] is usually referred to as reactance in the electromagnetism/microwave cavity literature~\cite{ZhengOtt06}. 
The semiclassical autocorrelation function $A_p(t)$ is finally all we need to approximate the envelopes of both local density of states and reactance.
\begin{figure}[tbh!]
\centerline{
(a)
\includegraphics[width=7.8cm]{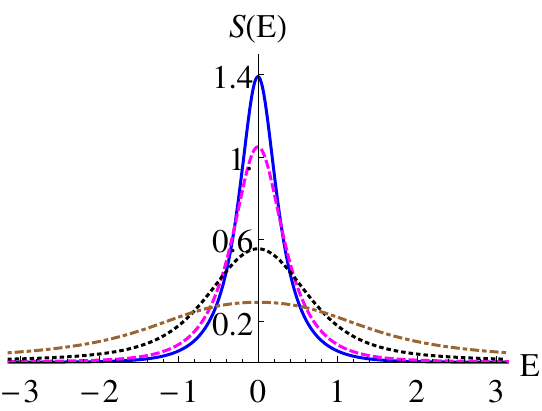}
(b)
\includegraphics[width=7.8cm]{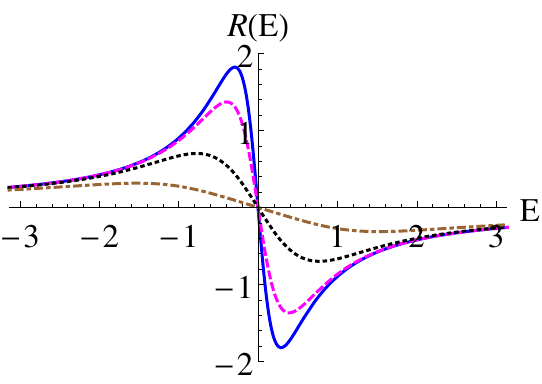}
} 
\centerline{
(c)
\includegraphics[width=7.8cm]{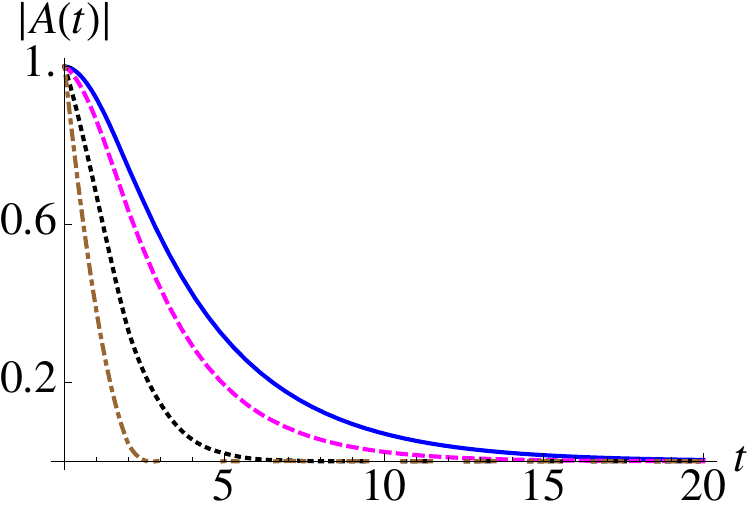}
}     
\caption{Semiclassical predictions~(\ref{ubersimpS}) for (a) the local density of states and (b) the reactance (real part of the same equation), where $\langle x|G_0(E)|x\rangle=-R_0(E)+iS_0(E)$, and $R_0(E)$ and $S_0(E)$ are approximated by the semiclassical envelopes in Eq.~(\ref{reacldos}), with the autocorrelation function given by Eq.~(\ref{ScarA}).
 Here $|x\rangle=|a\rangle$. Solid line: closed system. Dashed line: open system, $\Gamma=0.1$. Dotted line: open system, $\Gamma=0.5$; Dot-dashed line: open system, $\Gamma=1.5$ .
 (c) Corresponding autocorrelation functions, evaluated as Fourier transforms of the local densities of states.}
\label{scarpreds}
\end{figure}
      
Real and imaginary parts of Eq.~(\ref{reacldos}), that together make the spectral density of the closed system, are thus plugged into Eq.~(\ref{ubersimpS}) for the spectral density of the open system. To have an idea of the outcome, we first explore the qualitative features of this approximation on the simple model of a fixed point with orthogonal stable/unstable manifolds [cf. Eq.~(\ref{ScarA})]. As shown in Fig.~\ref{scarpreds}(a), the local density of states, which is peaked when the system is closed, gradually  flattens by increasing dissipation, and tends to a uniform distribution in the limit of large couplings with the opening. That means that the very signature of localization fades away, and it appears as though dissipation suppresses the system-specific effects of scars, and restores the properties of RMT. One can immediately think of the dynamical implications of this result, first of all concerning the autocorrelation function, which is now expected to decay faster with an increased coupling to the opening, toward the RMT scenario [Fig.~\ref{scarpreds}(c)]. In the next section we illustrate how this phenomenon seemingly contradicts the loss-induced enhancement of localization observed in the Husimi distributions, and in the statistics of wavefunctions. It is also shown how dissipation restores symmetry in a skewed local density of states, and thus voids the effect of the non-orthogonality of the stable/unstable manifolds.             
  
\section{Numerical tests}
\label{numer}
 The above predictions are tested using quantum maps. Two models are considered in what follows, with different time-reversal symmetries. First, the quantized cat map is examined.
 Computation of eigenfunctions and local density of states of this model without dissipation were already introduced in Figs.~\ref{closednoscar} and~\ref{closedscar}. The classical evolution of the cat map is given by~\cite{creagh,Cr_Lee} 
 \begin{equation}
F_\epsilon = F_0 \circ M_\epsilon,
\label{eq:cat_map}
\end{equation}
with
\begin{equation}
\fl
F_0 : \left( \begin{array}{cc}
q' \\ p'
\end{array} \right) =
\left( \begin{array}{cc}
1 & 1 \\
1 & 2
\end{array} \right)
\left( \begin{array}{cc}
q \\ p
\end{array} \right)  \hspace{0.3cm} \mathrm{mod} 1
\, ,
\hspace{0.5cm}
M_\epsilon :
\left( \begin{array}{cc}
q' \\ p'
\end{array} \right) =
\left( \begin{array}{cc}
q - \epsilon\sin(2\pi p) \\
p  \\
\end{array} \right) \hspace{0.3cm} \mathrm{mod} 1
\label{eq:class_cat}
\end{equation}
$F_\epsilon$ has an inversion symmetry about the origin, plus the anticanonical time-reversal symmetry
\begin{equation}
PF_\epsilon P = F^{-1}_\epsilon
\, ,
\label{PFPTRS}
\end{equation}
with
\begin{equation}
P \left( \begin{array}{cc}
q \\ p
\end{array} \right) =
\left( \begin{array}{cc}
-1 & 0 \\
-1 & 1
\end{array} \right)
\left( \begin{array}{cc}
q \\ p
\end{array} \right)
\, ,
\label{Pqp}
\end{equation}
by which, in principle, the spectrum should follow COE statistics~\cite{Cr_Lee,KeatMezz}.
The quantization of the map is given by \cite{creagh,han_ber}
\begin{equation}
U_\epsilon = U_0V_\epsilon
\label{eq:ueps}
\end{equation}  
where
\begin{eqnarray}
\nonumber
\left<q_j|U_0|q_k\right> &=& N^{-1/2}\rme^{i\pi/4}
\rme^{2\pi Ni(q_j^2-q_jq_k+q_k^2/2)} 
\, ,
\\
\left<q_j|V_\epsilon|q_k\right> &=& \sum_{p_m}
\frac{1}{N}\rme^{Ni\left(-\epsilon\cos2\pi p_m
+2\pi(q_j-q_k)p_m\right)}
\, .
\label{eq:veps}
\end{eqnarray}
The amplitude of the nonlinear perturbation of the cat map is henceforth set to $\epsilon=0.1$. 
The $|a\rangle$ identifying the opening is a minimum-uncertainty Gaussian wavepacket $|q_0,p_0\rangle$ already introduced in Eq.~(\ref{wpack}).  A non-Hermitian Hamiltonian of the form~(\ref{HnonH}) results in a non-unitary quantum propagator, which is here realized as~\footnote{The dissipative part of the propagator is implemented as a power series, which is truncated when sufficient convergence has been achieved.}   
\begin{equation}
U =  e^{-\Gamma\left|a\right>\left<a\right|}U_0
\,.
\label{eq:incl_leaks}  
\end{equation}  
\begin{figure}[tbh!]
\centerline{
(a)
\includegraphics[width=7.cm]{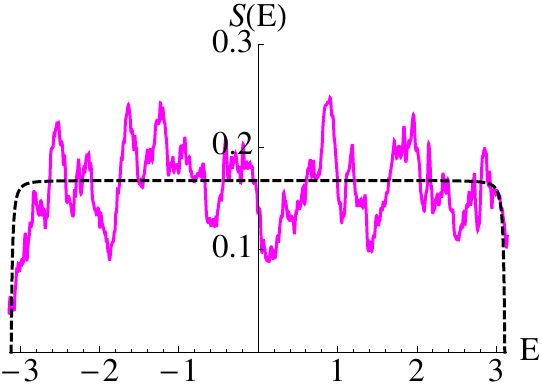}
(b)
\includegraphics[width=7.cm]{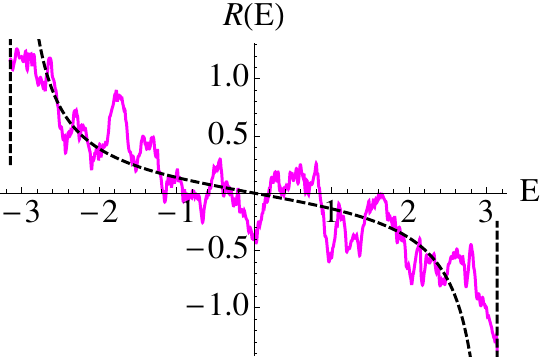}
}  
\caption{(solid line) The numerically evaluated local density of states~(\ref{opnldosreact}) for the fully chaotic perturbed cat map~(\ref{eq:veps}) with (a) $|a\rangle=|x\rangle$ a coherent state centered at a random point in phase space away from short periodic orbits and thus with no scarring effects; (dashed line) the mean-field estimate~(\ref{avG}) where the COE average~(\ref{gCOE}) is used for the expectation value of the spectral density of the closed system; (b) the reactance $R(E)$ computed from Eq.~(\ref{opnldosreact}) from the same simulation, with the relative RMT estimate (dashed line).}     
\label{numldosopn}
\end{figure}    
A sample of 2310 states is examined by numerical diagonalization of 20 realizations of the matrix~(\ref{eq:incl_leaks}), whose dimension is varied from 200 to 220 (even numbers).
Local density of states and reactance are then computed from Eq.~(\ref{opnldos}):
\begin{equation}
\fl
 S(E) = \mathrm{Re} \sum_n h_n(x)\frac{\gamma_n+i(E-\varepsilon_n)}{\gamma_n^2+(E-\varepsilon_n)^2} 
\,,
\hspace{1cm}
R(E) = \mathrm{Im} \sum_n h_n(x)\frac{\gamma_n+i(E-\varepsilon_n)}{\gamma_n^2+(E-\varepsilon_n)^2}
\,,  
\label{opnldosreact}
\end{equation}
where we recall that $h_n(x)=  \frac{\langle x|\Psi_n\rangle\langle\Phi_n|x\rangle}{\langle\Phi_n|\Psi_n\rangle}$.    
Figure~\ref{numldosopn} shows local spectrum and reactance of the quantum cat map opened through a single channel. Importantly, a non-scarred state is chosen as probe state and open channel (here, as in Section~\ref{1cpreds}, $|x\rangle=|a\rangle$), so that the local spectrum shows no notable deviations from the uniform distribution one should expect for the closed system, whose spectral statistics follows the Circular Orthogonal Ensemble (COE). 
For comparison to the theoretical expectations derived in the previous section, $S(E)$ and $R(E)$ are regarded as  real and imaginary parts of the expectation value of the Green's function (spectral density), $\langle x|G(E)|x\rangle$. 
Opening $|a\rangle$ and probe state $|x\rangle$ coincide, so that the predicted local density of states should be given by Eq.~(\ref{ubersimpS}). Since opening and probe state have been placed away from short periodic orbits, the RMT ensemble averages can be taken, so that the result~(\ref{avG}) applies for local density of states and reactance.
Figure~\ref{numldosopn} indeed confirms this simple prediction.     

Next, scarring is considered: the wavepackets representing opening and probe states are both centered at the origin, where the map~(\ref{eq:cat_map}) has a fixed point. Here, the 
autocorrelation function $A(t)$ is approximated by the semiclassical expression~\cite{Kap99,Cr_Lee}
\begin{equation}
A_p(t) = \frac{\rme^{i\arctan{\frac{Q\sinh{\lambda t}}{2\cosh{\lambda t}}}}}
{\left(\cosh^2{\lambda t}+Q^2\sinh^2{\lambda t}\right)^{1/4}}
\,,
\label{CatAp}
\end{equation}
where $Q$ is the angle between stable and unstable manifolds, while $\lambda$ is the stability exponent of the
fixed point of the perturbed cat map. The full derivation of Eq.~(\ref{CatAp}) is available in Appendix A. 
Equation~(\ref{CatAp}) is a generalization of the semiclassical estimate~(\ref{ScarA}) for non-orthogonal stable/unstable manifolds. As shown in Fig.~\ref{closedscar}, both $A_p(t)$ and its Fourier  transform $S_p(E)$ perfectly match the autocorrelation function and local density of states of the closed system. 
\begin{figure}[tbh!]
\centerline{
(a)
\includegraphics[width=6cm]{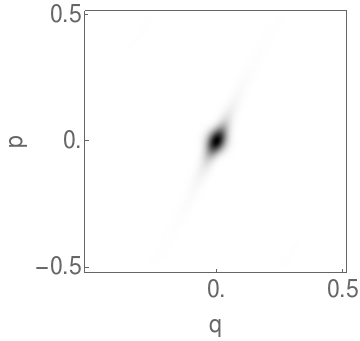}
(b)
\includegraphics[width=6cm]{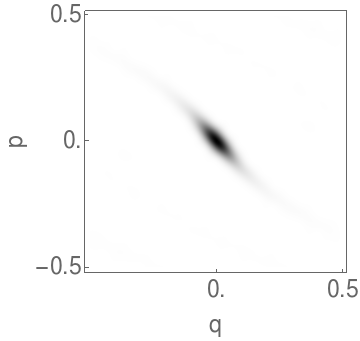}
}
\centerline{
(c)
\includegraphics[width=8cm]{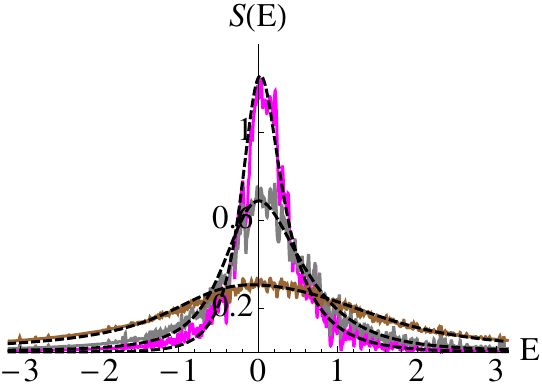}
(d)
\includegraphics[width=8cm]{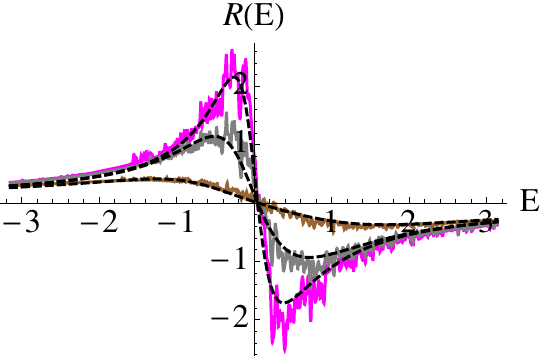}
}
\centerline{
(e) 
\includegraphics[width=8cm]{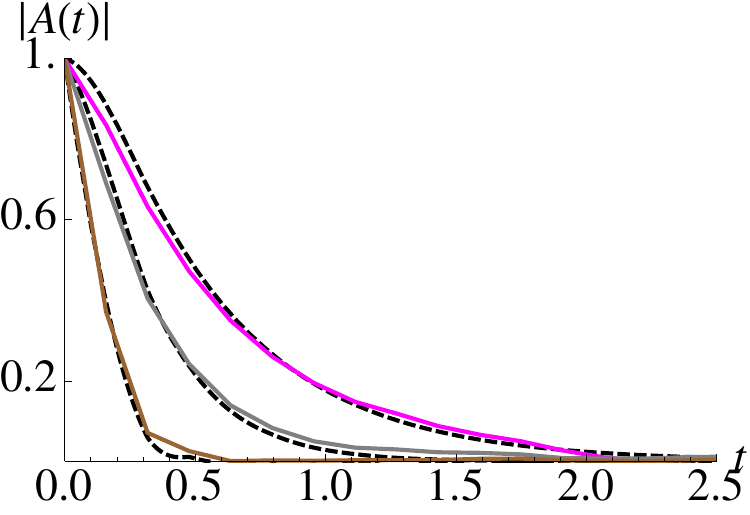}
}
 \caption{(a) The Husimi projection of a left scarred eigenfunction of the non-unitary propagator~(\ref{eq:incl_leaks}) for the quantum cat map, with $\Gamma=0.05$. (b) The projection of the corresponding right eigenstate. 
(c) Local density of states of the same system with $\Gamma=0.05$ (tallest peak), $\Gamma=0.07$ (middle peak), $\Gamma=0.15$ (flattest curve). (d) Reactance for the same simulation, with the coupling parameters in the same order from the tallest peak/ deepest trough to the lowest peak/shallowest trough. Solid lines: numerics; dashed lines: theory [Eq.~(\ref{reacldos})]. (e) Corresponding autocorrelation functions, evaluated as Fourier transforms of the local densities of states.}        
\label{catopnldos}
\end{figure}
As for the open system, 
both local density of states and reactance are estimated from Eq.~(\ref{ubersimpS}). In that expression, the expectation value of the Green's function of the closed system $\langle x|G_0(E)|x\rangle$ is evaluated by means of Eq.~(\ref{reacldos}), where the crucial autocorrelation function $A(t)$ is in turn replaced by the semiclassical $A_p(t)$ above. Figure~\ref{catopnldos}(c)-(d) shows that the numerically evaluated $S(E)$ and $R(E)$ closely follow their semiclassical expectations. In the case of weak to moderate coupling to the opening [$\Gamma\lesssim 0.05$, Fig.~\ref{catopnldos}(c)-(d)], the peak of the normalized local density of states gradually lowers, and the agreement between theory and numerics is quantitative with no fitting parameters. The numerically computed reactance, that has no normalization, matches the semiclassical estimate up to a multiplicative constant.   
For large dissipation [$\Gamma>0.05$, Fig.~\ref{catopnldos}(c)-(d)],       
one can indeed observe the predicted flattening of the peak, as the envelope of the local density of states gradually returns to the RMT-predicted uniform distribution. Remarkably, the asymmetry of the envelope, originally due to the non-orthogonality of the stable/unstable manifolds, is also suppressed by dissipation.  
The agreement visible in this regime [e.g. bottom peak of Fig.~\ref{catopnldos}(c)]  is qualitative, since, as in the single-channel theory, the coupling $\Gamma$ is no longer 
the same for theory and numerics, and hence becomes a fitting parameter. This shortcoming is ascribed to the nature of the mean-field theory, that
implies an assumption of `collectivization of widths', as phrased in references~\cite{SokZel89,SokZel92}. In the present context, that hypothesis is tantamount to confuse the
control parameter $\Gamma$ with the order parameter, probably $\langle x|W|\psi_n\rangle$, which evidently depends on the single wavefunction, and, importantly, does not respond linearly to changes in the coupling parameter $\Gamma$, except in perturbative regime. 
Having said that, the quantitative accuracy of the present semiclassical prediction breaks down at relatively large couplings, and the theory otherwise proves reliable.          

The progressive shift from an energy-dependent, sharply peaked local density of states to a uniform distribution is due to the losses that cause spectral linewidths to widen significantly at large dissipation. In appearance, though, this behavior is in contrast with that of the Husimi projections of the scarred eigenstates, where, instead, localization seems to be enhanced by the opening [Fig.~\ref{catopnldos}(a)-(b)]. In reality, the latter behavior only concerns the eigenstates that exhibit scarring, and the Husimi distributions do not yield information on dissipation, namely the width or decay rate of the corresponding resonances. On the other hand, the spectral density depends on all the resonances, and can describe the impact of dissipation on the spectrum and the dynamics.
The phenomenology observed here 
can be interpreted using a mode-mode interaction picture: 
as dissipation increases, the few scarred eigenstates (`doorway states')~\cite{SokZel97,RichSScars} become localized and lossy, thus smoothening the local spectrum, 
while they separate from a multitude of long-lived eigenstates (`trapped resonances')~\cite{SokZel89,ISSO94,rotlet}.


\section{Multiple-channel opening}
\label{mchan}
As one of the sources of inspiration for this work is the alleged scar enhancement in the numerical  simulation of dielectric microcavities at optical frequencies, the model considered in what follows consists of a time-reversal symmetric quantum map, and a partial, projective opening onto a subset of the phase space,
defined by Fresnel laws of refraction for a given refractive index. The theory for a multichannel opening is based on the expression~(\ref{Dyson}), and additional mean-field approximations will be made 
in the following sections, depending on the relative location of opening and scar/probe state.  

\begin{figure}[tbh!]
\centerline{
(a)
\includegraphics[width=6.5cm]{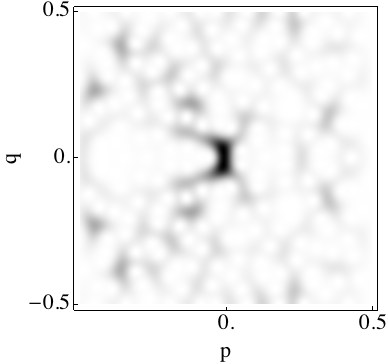}
(b)
\includegraphics[width=7.5cm]{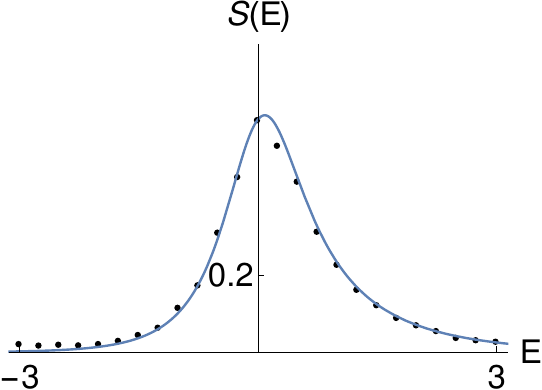}
}  
\caption{(a) Husimi distribution of a scarred eigenstate of the quantum map~(\ref{kicked_opt}).
(b) Normalized histogram (dots) versus theoretical prediction~(\ref{CatAp}) (solid line) for the local density of states of the quantum map~(\ref{kicked_opt}). }        
\label{OpticLdosCl}
\end{figure}

\begin{figure}[tbh!]
\centerline{
(a)
\includegraphics[width=6.5cm]{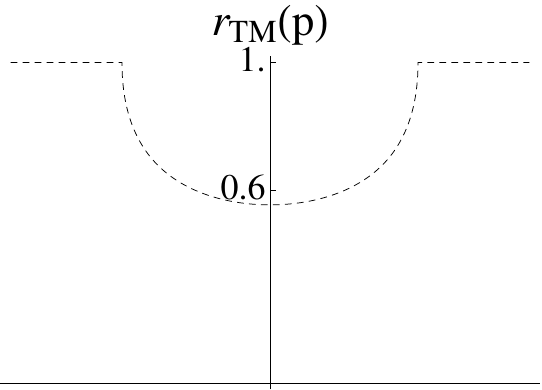}
(b)
\includegraphics[width=6.5cm]{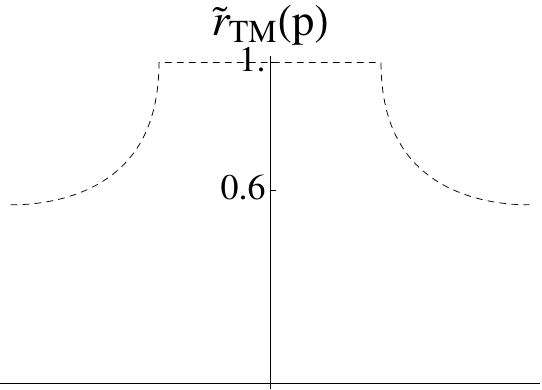}
}  
\caption{(a) The reflection coefficient yielding the Fresnel-type of opening~(\ref{tm_r}), vs. momentum $p$ for refractive index $n_r=3.5$ .  
(b) The `swapped' reflection coefficient $\tilde{r}_{\mathrm{TM}}$ that opens the external stripes of the phase space.}        
\label{r_openings}
\end{figure}

For the numerical tests, we choose the transformation    
\begin{equation}
M_\kappa = M_0 \circ f_\kappa,
\label{eq:cat_map}
\end{equation} 
with
\begin{equation}
\fl
M_0 : \left( \begin{array}{cc}
p' \\ q'
\end{array} \right) =
\left( \begin{array}{cc}
2 & 1 \\
3 & 2
\end{array} \right)
\left( \begin{array}{cc}
p \\ q
\end{array} \right)  \hspace{0.3cm} \mathrm{mod} 1
\, ,
\hspace{0.3cm}
f_\kappa :
\left( \begin{array}{cc}
p' \\ q'
\end{array} \right) =
\left( \begin{array}{cc}
p \\
q + \kappa\sin(2\pi p) \\
\end{array} \right) \hspace{0.3cm} \mathrm{mod} 1
\,.
\label{eq:class_trev_cat}
\end{equation}
The nonlinear kick is here dynamically inessential as the linear map on a torus is already fully chaotic, but it importantly rids the quantized model
of some undesirable symmetries due to number theoretical properties, that would bias the statistics of its spectrum~\cite{KeatMezz}. For that reason, we will
set $\kappa=10^{-5}$ in the quantum computations. The kicked map $M_\kappa$ instead bears 
no symmetry but the time-reversal one, which, unlike for the map~(\ref{eq:class_cat}), is \textit{canonical}, that is
\begin{equation}
TM_\kappa T = M^{-1}_\kappa
\, ,
\label{TrevSymm}
\end{equation}
with
\begin{equation}
T \left( \begin{array}{cc}
p \\ q
\end{array} \right) =
\left( \begin{array}{rr}
1 & 0 \\
0 & -1
\end{array} \right)
\left( \begin{array}{cc}
p \\ q
\end{array} \right)
\,.
\label{Tpq}
\end{equation}
The canonical time-reversal symmetry produces real-valued eigenfunctions in the following quantized version of the map, made of 
the linear hyperbolic shear $M_0$~\cite{han_ber}
\begin{equation}
\left<p_j|U_0|p_k\right> = N^{-1/2}\rme^{i\pi/4}
\rme^{2\pi Ni(p_j^2-p_jp_k+p_k^2)} 
\,,
\label{hbquant}
\end{equation}
as well as the kick $f_\kappa$:
\begin{equation}
\left<p_j|V_\kappa|p_k\right> = \rme^{-i N\kappa\cos(2\pi p_j)} 
\,,
\label{trkick}
\end{equation}
or
\begin{equation}
\left<p_j|V_\kappa|p_k\right> = \rme^{-i N\kappa\cos(2\pi p_k)} 
\,.
\label{trkick2}
\end{equation}
For that reason the whole map is quantized by giving two half-kicks to each momentum of the matrix element, with the linear hyperbolic shear in between, in the following way:
\begin{equation}
\langle p_j|U| p_k\rangle = \langle p_j|V_{\kappa/2}U_0V_{\kappa/2}|p_k\rangle
\,.
\label{kicked_opt}
\end{equation}
Just like the previous quantum cat map~(\ref{eq:veps}), Eq.~(\ref{eq:class_trev_cat})  has a fixed point at the origin and the quantization~(\ref{kicked_opt}) bears a visible scar at the same location [Fig.~\ref{OpticLdosCl}(a)].
The envelope of the local density of states is calculated for the closed system via the Fourier transform of the semiclassical formula~(\ref{CatAp}) for the autocorrelation function
 already seen in the single-channel theory.
The agreement with the numerics is shown in figure~\ref{OpticLdosCl}(b).
The multiple-channel opening is realized in a similar way to a microresonator~\cite{SchomKeat} for transverse-magnetic (TM) wave propagation, that is by considering the cat map as a surface of section of the dynamics at the boundaries, with reflection coefficients
\begin{equation}
r_{\mathrm{TM}} = \left(\frac{\sqrt{1-p^2}-\sqrt{n_r^{-2}-p^2}}{\sqrt{1+p^2}+\sqrt{n_r^{-2}-p^2}}\right)^2
\,,
\label{rtm}
\end{equation}
where $n_r$ is the chosen refraction coefficient.
 Quantizing under the assumption that momentum is conserved at each iteration, we obtain  
 the subunitary projection operator 
 \begin{equation}
R = \sum_jr(p_j)|p_j\rangle\langle p_j|
\, , 
\label{Ropn}
\end{equation}
where 
\begin{equation}
r(p) = \left\{ \begin{array}{cc}
1  &  |p|>\frac{1}{n_r} \\
\\ 
\sqrt{r_{\mathrm{TM}}}(p) &  |p|\leq\frac{1}{n_r}
\end{array} \right. 
\,,
\label{tm_r}
\end{equation}
or, alternatively, 
\begin{equation}
\tilde{r}(p)=r\left(\frac{1}{2}-|p|\right)
\,,
\label{side_r}
\end{equation} 
to leave the central region of the phase space closed and, instead, open the
outer stripes. Both types of openings are pictured in Fig.~\ref{r_openings}.

Then, the nonunitary propagation is simply $U_R=RU$, and it may mimic
 the dynamics inside a chaotic dielectric microcavity.    
Now, in a continuous-time representation, the corresponding non-Hermitian Hamiltonian would be expressed as 
\begin{equation}
H = H_0 + i\sum_{j}\ln r(p_j)|p_{j}\rangle\langle p_{j}|
\,.
\label{FresnH}
\end{equation}
 Then (assume $\hbar=1$, take $t=1$) 
 \begin{eqnarray}
 \fl
 \nonumber
 U_R = e^{-iH} =e^{\sum_{j}\ln r(p_j)|p_{j}\rangle\langle p_{j}|} e^{-iH_0} =
   \left[\mathds{1} + \sum_{j}\ln r(p_j)|p_{j}\rangle\langle p_{j}| + \frac{1}{2}\left( \sum_{j}\ln r(p_j)|p_{j}\rangle\langle p_{j}|\right)^2 + ... \right]e^{-iH_0}   \\
   \fl =
 \left[\sum_j \e^{\ln r(p_j)}|p_j\rangle\langle p_j|\right]U = RU
 \,,
 \end{eqnarray}
where we have used the idempotent and orthogonal properties of the projectors $|p_j\rangle\langle p_j|$.
\subsection{Opening away from the scar}
\label{scaraway}
The case is first considered where the opening does not enclose the probe state $|x\rangle$ and thus the scar, 
meaning that the outer stripes of the phase space (the unit torus) are open [Fig.~\ref{r_openings}(b)], while the scar at the origin is
in the closed region. This is given mathematically by the expression~(\ref{side_r}).
We now obtain a prediction for the spectral density of the open system, $\langle x|G(E)|x\rangle$, based on both scar theory and RMT.  
The strategy is that to start from the solution~(\ref{Dyson}) to the Dyson equation, and to separate the expectation  $\langle x|G_0(E)|x\rangle$,
approximated with the semiclassical estimate $G_{\mathrm{sc}}(E,x_p)$, from $\langle p_j|G_0(E)|p_j\rangle$,
assumed to be random, and, as such, averaged over the appropriate RMT ensemble. 
We now use a mean-field approximation to estimate the spectral density, starting from the exact result
\begin{equation}
\fl
\langle x|G(E)|x\rangle =
\langle x|G_0(E)|x\rangle -  i\,\langle x|G_0(E)W \left[1 + i\,G_0(E)W\right]^{-1} G_0(E)|x\rangle
\,.
\label{mc_spctdens}
\end{equation} 
As seen in the single-channel theory, $\langle x|G_0(E)|x\rangle\simeq  G_{\mathrm{sc}}(E,x)$,  given by the semiclassical theory, while we focus on the
second term of Eq.~(\ref{mc_spctdens}). The idea is to average over the ensemble, and apply a mean-field approximation, so that
\begin{equation}
\fl
\overline{\langle x|G_0(E)W \left[1 + i\,G_0(E)W\right]^{-1} G_0(E)|x\rangle} \simeq 
\frac{\overline{\langle x|G_0(E)WG_0(E)|x\rangle}}{\overline{1 + i\,G_0(E)W}}
\,,
\label{meanfield}
\end{equation}
and thus we now evaluate the ensemble averages of numerator and denominator separately:
\begin{equation}
\fl
\overline{1 + i\,G_0(E)W} = 1 + i\,\overline{\tr G_0(E)W} = 1 + i\sum_j \ln r(p_j)\overline{\langle p_j|G_0(E)|p_j\rangle} =
1 + i\frac{w}{N}g_0(E)
\,,
\label{denom}
\end{equation}
where we have called $w=\sum_j \ln r(p_j)$ for convenience, while $g_0(E)=\overline{\langle p_j|G_0(E)|p_j\rangle}$, and
in this case it is estimated by means of the COE result~(\ref{gCOE}), because none of the $p_j$ in the open region overlap with the scar.
On the other hand, the numerator is averaged over the RMT ensemble as
\begin{eqnarray}
\fl
\nonumber
\overline{\langle x|G_0(E)WG_0(E)|x\rangle} = \sum_j \ln r(p_j)\overline{\langle x|G_0(E)|p_j\rangle\langle p_j|G_0(E)|x\rangle} = \\ 
\fl
 =  \sum_j \ln r(p_j)\sum_{m,n}^N\overline{\langle x|\psi_n\rangle\langle \psi_n|G_0(E)|\psi_n\rangle\langle\psi_n|p_j\rangle\langle p_j|\psi_m\rangle
  \langle \psi_m|G_0(E)|\psi_m\rangle\langle\psi_m|x\rangle}
\,.
\label{num_one}
\end{eqnarray}
Now, one can assume that the amplitudes $\langle p_j|\psi_n\rangle$ are random waves, uncorrelated from the other factors of Eq.~(\ref{num_one}).
Therefore, the sole diagonal contributions ($n=m$) survive ensemble averaging in the previous expression, which we may consequently rewrite as
\begin{eqnarray}
\nonumber
\fl
\overline{\langle x|G_0(E)WG_0(E)|x\rangle} &=&  \sum_j  \ln r(p_j)\sum_{n}^N\overline{\left(|\langle x|\psi_n\rangle|^2\,\langle \psi_n|G_0(E)|\psi_n\rangle\right)
 \left(|\langle p_j|\psi_n\rangle|^2\,\langle \psi_n|G_0(E)|\psi_n\rangle\right)} \\
 &=& \sum_j  \ln r(p_j)\overline{\langle x|G_0(E)|x\rangle\,\langle p_j|G_0(E)|p_j\rangle}
\,,
\label{num_two}
\end{eqnarray}
and, consistently with the physical picture of the system at hand (that is scar at $|x\rangle$, while randomness everywhere else), we take the 
ensemble average of the spectral density of each open channel, while we stick to the semiclassical approximation for $\langle x|G_0(E)|x\rangle$:
\begin{equation}
\fl
\overline{\langle x|G_0(E)WG_0(E)|x\rangle} \simeq \sum_j \ln r(p_j)\overline{\langle p_j|G_0(E)|p_j\rangle}\,G_{\mathrm{sc}}(E,x) =
\frac{w}{N}g_0(E)\,G_{\mathrm{sc}}(E,x)
\,.
\label{num_three}
\end{equation}      
At this point, we may put the pieces together to estimate the spectral density~(\ref{mc_spctdens}) as
\begin{equation}
\fl
\overline{\langle x|G_0(E)|x\rangle} = G_{\mathrm{sc}}(E,x) - i\frac{w}{N}\frac{g_0(E)\,G_{\mathrm{sc}}(E,x)}{1 + i\frac{w}{N}g_0(E)}
=  \frac{G_{\mathrm{sc}}(E,x)}{1 + i\frac{w}{N}g_0(E)} 
\,.
\label{sideopen_est}
\end{equation}
An observation is in order: in separating averages and discarding off-diagonal contributions from Eq.~(\ref{num_one}), it is central to have $\langle x|p_j\rangle=0$.
That is, none of the open channels, located at the sides of the unit torus in the phase space, overlap with the probe state, that we instead still choose at the scar and 
therefore around the center of the torus. 
\begin{figure}[tbh!]
\centerline{
(a)
\includegraphics[width=7.5cm]{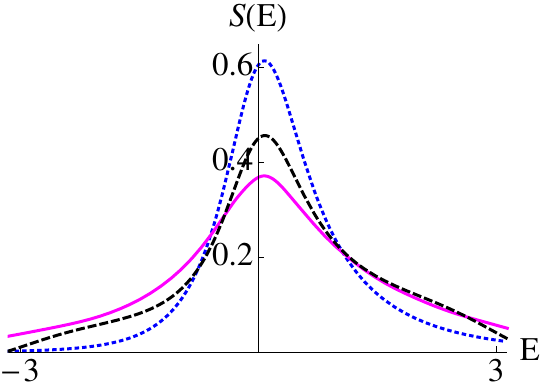}
(b)
\includegraphics[width=7.5cm]{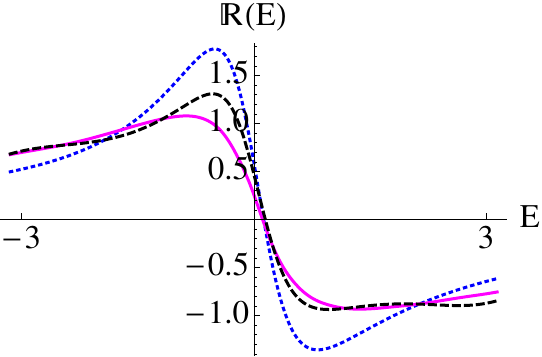}
}   
\centerline{
(c)
\includegraphics[width=7.5cm]{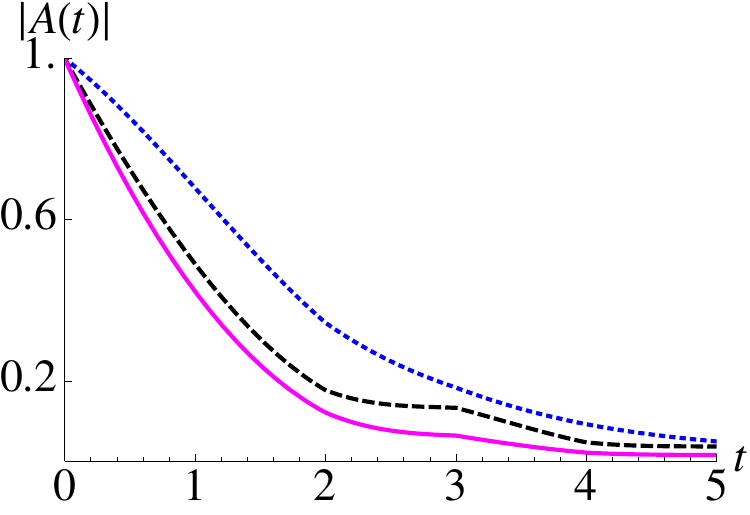}
}
\caption{Local density of states (a) and reactance (b) with the Fresnel two-strip opening given by Eq.~(\ref{side_r}) ($p_j>|1/n_r|$), with $n_r=2.5$. Solid line: numerics;  dashed line: mean-field prediction~(\ref{sideopen_est}); dotted line: semiclassical estimate for the closed system, included for comparison. (c) Corresponding autocorrelation functions, evaluated as Fourier transforms of the local densities of states. }        
\label{aneqx}
\end{figure}    

Figure~\ref{aneqx} shows an example of how the mean-field estimate~(\ref{sideopen_est}) fits numerical evaluations of local density of states and reactance for the quantized time-reversal symmetric 
cat map, coupled to the opening~(\ref{side_r}).  The agreement of the theory with the numerics appears adequate, as, importantly, there are no fitting parameters. 
In particular, the mean-field theory formulated here well captures the qualitative behavior of the spectral density. Specifically in
the present case of a `classical' opening placed away from the scar at the origin, the local density of states (Fig.~\ref{aneqx}(a), plotted together with the local spectrum of the closed system for reference) flattens as in the single-channel, coherent-state opening, but notably, the characteristic peak 
remains, while the tails of the distribution rise considerably. As summarized in Eq.~(\ref{sideopen_est}), the interaction of random- and scarred states is most evident in this model. On the other hand,
the reactance [Fig.~\ref{aneqx}(b)] does not deviate considerably from that of the closed system in the range of $E$ occupied by the scarred states, while it tends to become smoother, almost energy-independent, outside that range.

\subsection{Opening overlapping with the scar}
\label{inscar}
We now consider an opening, still inspired by optics, that opens the middle stripe of the phase space, this time enclosing the scar at the origin, that we have been probing, according
to the expression~(\ref{tm_r}).
Looking again for a system-independent mean-field theory that predict the salient features of the spectral density, we notice that the treatment of the previous section may not 
apply to the present opening $\sum_jr(p_j)|p_j\rangle\langle p_j|$, which now \textit{encloses} the scar-centered probe state $|x\rangle$, and thus $\langle p_j|x\rangle\neq0$ .
That prevents us in principle from separating the ensemble averages $\overline{\langle x|G_0(E)|x\rangle\,\langle p_j|G_0(E)|p_j\rangle}$, since the two spectral densities are now
correlated. We will thus opt for an alternative approach to mean-field theory, involving  
a $P-Q$ formalism in the spirit of Feshbach's work on nuclear reactions~\cite{Fesh}: 
define two projection operators $P=|x\rangle\langle x|$, and $Q=\mathds{1}-|x\rangle\langle x|$.
Letting $|x\rangle$ be the probe state, again a wavepacket centered at the fixed point of the 
map, we may now restrict the analysis to the projected Hamiltonian $PHP$, that is
\begin{equation}
\fl
PHP = |x\rangle\langle x|H_0|x\rangle\langle x| +i\sum_j \ln r(p_j) |x\rangle|\langle x|p_j\rangle|^2\langle x| = h_0 |x\rangle\langle x| - i\tilde{\Gamma}(x) |x\rangle\langle x|
\,,
\label{effham}
\end{equation} 
where $h_0=\langle x|H_0|x\rangle$, while $\tilde{\Gamma}(x)=-\sum_j \ln r(p_j)|\langle x|p_j\rangle|^2$ is the effective coupling. At this point, the problem is recast into the single-channel non-Hermitian Hamiltonian~(\ref{singcH}), whose local Green's function is written as
\begin{equation}
G(E,x) = \frac{\langle x|G_0(E)|x\rangle}{1+i\tilde{\Gamma}(x)\langle x|G_0(E)|x\rangle}
\,,
\label{Geff}
\end{equation} 
and again the spectral density of the closed system is approximated with its semiclassical expression, $\langle x|G_0(E)|x\rangle\approx G_{\mathrm{sc}}(E,x)$. 
\begin{figure}[tbh!]
\centerline{
(a)
\includegraphics[width=8.cm]{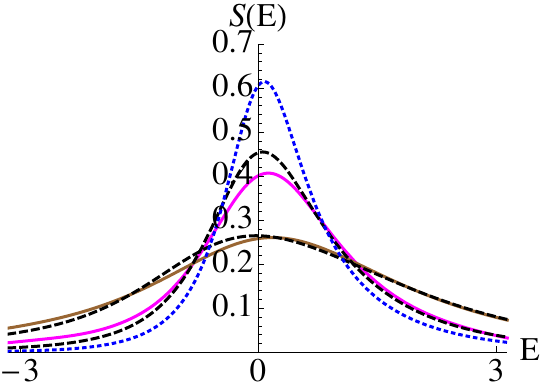}
(b)
\includegraphics[width=8.cm]{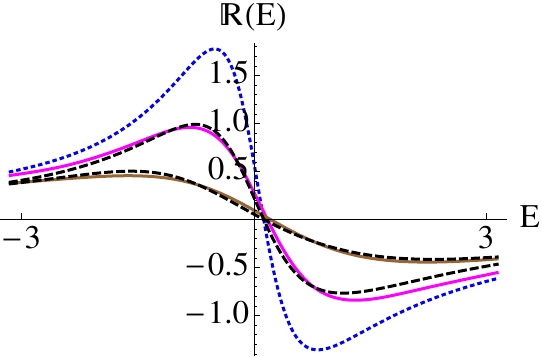}
} 
\centerline{
(c)
\includegraphics[width=7.5cm]{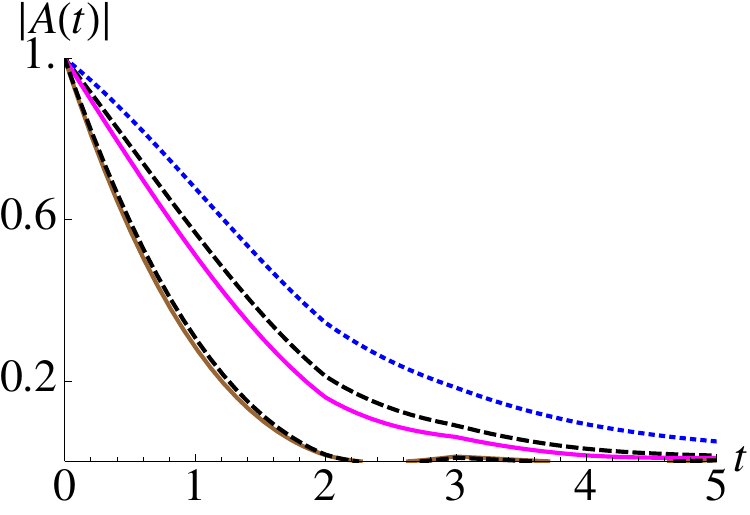}
}   
\caption{The numerically evaluated (a) local density of states and (b) reactance of the quantum map~(\ref{hbquant}) versus the mean-field expectations. Solid lines: numerical simulation of the  open system with (magenta) refractive index $n=2.5$,  and with (brown) $n=1.5$.  Dashed lines: predictions of the effective theory, Eq.~(\ref{Geff}). Dotted line: semiclassical prediction for the
closed system. (c) Corresponding autocorrelation functions, evaluated as Fourier transforms of the local densities of states.}        
\label{OpticLdos}
\end{figure} 
The results in this case are in all similar to what observed with the single-channel opening placed at the scar, except that the ensemble numerical computations yield a smoother spectral density.
That is due to the multichannel opening, that is less localized and produces resonances all of finite width, as opposed to the plethora of resonances `trapped' arbitrarily close to the real axis that
arise from a localized single channel. The effective theory~(\ref{Geff}) works well as long as the refractive index allows for both an open- and a closed region of the phase space, in other
words for $\frac{1}{n_r}<\frac{1}{2}$, half the width of the torus. In the figure, the highest peak plotted with a solid (magenta) line is obtained from numerics with $n_r=2.5$.    
For larger refractive indices, the effective theory is validated only qualitatively, with $\tilde{\Gamma}(x)$ a fitting parameter. Once again it is noted the progressive smoothening of the spectral density with stronger couplings to the continuum, an indication that the dominant scar at the origin still rules, as long as it is enclosed by the opening.  

\section{Conclusions, implications, prospects}

The all-too-well established results of Random Matrix Theory (RMT) do not apply to systems with localization such as scarring, and therefore, in closed as
in open systems, one needs to develop alternative treatments, in order to predict observables and measurable quantities such as the spectral density.
While convincing efforts in this direction have been available in the realm of closed systems (Hermitian Hamiltonians), here we have presented a fair attempt
to obtain expectation values for the spectral density of an open chaotic system (non-Hermitian Hamiltonian), featuring scarred states, and thus a spectral statistics
deviating significantly from RMT. The approach used is based on semiclassical estimates of the autocorrelation function in the neighborhood of the scar,
which, in energy domain, are then plugged into a (Dyson) self-consistent equation to estimate the expectation of the Green's function for the 
open system. 
 
Although obtained by means of a non-perturbative calculation, our predictions for the spectral density somehow inherit the hypothesis
of `collectivization of widths' put forward for RMT-governed systems, and that, in the present display, proves quantitatively accurate only for weak to 
moderate couplings to the continuum. Otherwise, semiclassical and mean-field estimates worked out in both single- and multichannel theories still
provide a qualitatively accurate picture of the behavior of the spectral density, as the coupling with the opening grows stronger. In particular it is 
noted the `return to randomness' , that is the suppression of the characteristic peak of the local density of states of scars because of dissipation,
that apparently clashes with the common perception for example within the optics community that `scarring is enhanced' by openings.
It should also be remarked that the semiclassical expressions for the spectral density, based on the linearization of dynamics around scars,
are more accurate in the realm of open systems, because the `nonlinear' effects of long-time recurrences are suppressed by the escape. 

A theory yielding reliable predictions of the spectral density of a scattering Hamiltonian that features both random and localized states can have 
far-reaching implications, not last the currently much glamorous science of many-body scars~\cite{SerbRev21}. 
Born as a counterexample to the Eigenstates Thermalization Hypothesis (ETH), the detection of localized states within supposedly chaotic many-body 
Hamiltonians has developed into as science of its own. These still exceptional but physically significant states do bear analogies with their one-particle
counterparts~\cite{LukinMBSc19}, and have recently found meaning well beyond the issue of thermalization~\cite{Turn18}.  
 Thus, as the current problems in this field gradually 
saturate, the now plentiful many-body scarologists will probably turn their attentions to open systems, at some point. In the light of the convincing parallel between one-particle and
many-particle scars drawn by recent works, there may well be room for applying the present technology/results to many-body localization 
in dissipative systems. 
The biggest challenge in that direction is, as already for closed systems, to find the classical dynamics underlying the many-body quantum dynamics~\cite{AltHaake12}.
That task is often not quite straightforward, due to the fact that typical models that produce quantum many-body scars do not lend themselves to a mean-field treatment, 
in particular the spin chains have a `small on-site Hilbert space'~\cite{SerbRev21}.
However, recently developed state-space representations, such as time-dependent variational projections~\cite{SlowTherm20}, may be a promising tool
to recover a quantum-to-classical correspondence in non-thermal many-body systems, as well as to allow for semiclassical estimates of  the spectral density
with and without dissipation.  

\section{Acknowledgments}
The author is partially supported by NSF China - Grant No. 11750110416-1601190090.
The ensemble averaging techniques used throughout the manuscript have mostly been learnt reading reference~\cite{Haake}.
The author thanks Barbara Dietz for clarifying several aspects and essential technical points in those methods.
The author is also indebted with Yan Fyodorov for suggesting how to estimate the averaged spectral density for COE, Eq.~(\ref{gCOE}),
that, surprisingly, seems to be absent from the mainstream literature on RMT, nuclear scattering, or quantum chaos.
The author thanks Dmitri Savin for helpful discussions on the mean-field approximations leading to the `collectivization-of-widths'
hypothesis of Sokolov and Zelevinsky's, that is also used through the present work.
Many thanks also to Dmitri Ivanov for helping achieve a better understanding of how Feshbach's $P-Q$ formalism, originally used to describe nuclear reactions,
may result in an effective non-Hermitian Hamiltonian, and to Lev Kaplan for raising alerts on `possible singularities'  
in the evaluation of the averaged spectral density, in the earliest stages of this work. That warning led to the separate real- and imaginary
estimation of $G(E)$  as in Eq.~(\ref{reacldos}),
which is central to the whole treatment.

\section*{Appendix A: The autocorrelation functions}
\label{SCAp}
The following is the derivation of the semiclassical expression~(\ref{CatAp}) for the autocorrelation function of a Gaussian wavepacket propagating in the neighborhood of a fixed point with non-orthogonal stable/unstable manifolds. As seen in section~\ref{LDoS}, the classical dynamics is linearized near the fixed point. Without loss of generality, the phase space evolution may be written as~\cite{Cr_Lee}
 $x(t)=M(t)x(0)$, that is
\begin{equation}
\left( \begin{array}{cc}
q(t) \\ p(t)
\end{array} \right) =
\left( \begin{array}{cc}
\cosh\lambda t & e^\delta \sinh\lambda t \\
 e^{-\delta} \sinh\lambda t & \cosh\lambda t
\end{array} \right)
\left( \begin{array}{cc}
q(0) \\ p(0)
\end{array} \right)
\, ,
\label{InvOsc}
\end{equation}
where, as before, $\lambda$ is the stability exponent of the fixed point, while $\delta$ is related to the angle between stable and unstable manifolds in a way that will become clear at the end of the calculation.
In order to obtain the semiclassical evolution of a wavepacket centered at the fixed point $x_p$, we may use the Wigner propagator~\cite{Berry89,CrWhel99,OzoRev98} in the vicinity of $x_p$:  
\begin{eqnarray}
\nonumber
K(q,p,t) &=& \frac{e^{i\left[W_p(q,p,t)+(E-H_p(q,p,t))t\right]/\hbar+iS_pt/\hbar}}
{\sqrt{\mathrm{det}[M(t)+I]}}
\approx  \frac{e^{-ix^T\left[J[(M(t)-I)/(M(t)+I)\right] x/\hbar+iS_pt/\hbar}}
{\sqrt{\mathrm{det}[M(t)+I]}} \\
&=&
\frac{e^{-i(e^\delta q^2-e^{-\delta}p^2)\sinh\lambda t/(1+\cosh\lambda t)\hbar+iS_p/\hbar}}
{\sqrt{2(1+\cosh\lambda t)}}
\,,
\label{WigProp}
\end{eqnarray}
with $J=\left( \begin{array}{cc}
0 & 1 \\
 -1 & 0
\end{array} \right)\,,$  $W_p(q,p,t)=\oint pdq\,$, and $I$ the identity matrix. The autocorrelation function is then given by
the expectation value of the propagator in Wigner representation, that is
\begin{eqnarray}
\fl
\nonumber
A_p(t) &=& \int K(q,p,t)\psi(q,p,0)dqdp \propto
\int \frac{e^{-i(e^\delta q^2-e^{-\delta}p^2)\sinh\lambda t/(1+\cosh\lambda t)\hbar+iS_pt/\hbar}}
{\sqrt{2(1+\cosh\lambda t)}}\,e^{-q^2/\sigma^2-\sigma^2 p^2}dqdp  \\ \fl &=& 
\frac{e^{iS_pt/\hbar}}{\sqrt{\cosh\lambda t+iQ\sinh\lambda t}}
\,,
\label{skewmanA}
\end{eqnarray} 
where $Q=\sinh\delta$ is identified as the cotangent of the angle between stable and unstable manifolds. The previous expression is equivalent to Eq.~(\ref{CatAp}), 
which assumes $S_p=0$.

\section*{Appendix B: The average spectral density for COE}
The unitary propagator $U^t$, our starting point, acts in discrete time, and the spectral density is the expectation value of the
Laplace transform of $U^t$:
\begin{equation}
\fl
\langle x|G(z)|x\rangle = \langle x| \sum_t^\infty z^t\,U^t|x\rangle =
\lim_{\epsilon\rightarrow0}\sum_{t,n}|\langle x|\psi_n\rangle|^2 z^t\,e^{-it (\theta_n -i\epsilon)}  = 
\sum_{n=1}^N \frac{\e^{i\theta_n}}{\e^{i\theta_n}-z}|\langle x|\psi_n\rangle|^2
\,.
\label{Gtheta}
\end{equation}
Because everything here lives on the unit circle, let $z=\e^{i\theta}$, and also let us allow the `gauge freedom'
\begin{equation}
G(z) \longrightarrow i\,G(z)
\label{gaugeG}
\,.
\end{equation} 
Now, the ensemble average of the spectral density can be separated in intensities $\overline{|\langle x|\psi_n\rangle|^2}=\frac{1}{N}$, and 
propagator  
\begin{eqnarray}
\fl
\nonumber
\frac{1}{N}\langle \tr G(z) \rangle = \frac{i}{2\pi N}\,\sum_{n=1}^N\int d\theta_1\cdot\cdot\cdot d\theta_N \frac{\e^{i\theta_n}}{z-\e^{i\theta_n}}
|\e^{i\theta_1}-\e^{i\theta_2}|\,|\e^{i\theta_2}-\e^{i\theta_3}|\cdot\cdot\cdot|\e^{i\theta_{N-1}}-\e^{i\theta_N}| \\ \nonumber
= \frac{i}{2\pi N}\,\sum_{n=1}^N\oint dz_1\cdot\cdot\cdot dz_N \frac{z_n}{z-z_n}\frac{1}{iz_n}
|z_1-z_2|\,|z_2-z_3|\cdot\cdot\cdot|z_{N-1}-z_N| \\ 
= \frac{i}{2z} = \frac{1}{2}\sin\theta - \frac{i}{2}\cos\theta
\,,
\label{Gzavg}
\end{eqnarray}
where Plemelj formula~\cite{Kanwal97} was used in the evaluation of the loop integrals with a singularity along the integration path
\begin{equation}
\frac{1}{2\pi i}\oint_L \frac{g(\tau)}{\tau-z}d\tau = \frac{1}{2}g(z_0) +  \frac{1}{2\pi i}\, \mathrm{P}\,\oint_L \frac{g(\tau)}{\tau-z_0}d\tau
\label{Plemelj}
\,,
\end{equation}
and, importantly, $z_0\in L$.

The estimate~(\ref{Gzavg}) is consistent with the RMT result for the average density of states of COE, in fact
\begin{equation}
\rho(\theta)d\theta = -\frac{1}{\pi}\,\mathrm{Re}\, \left[\frac{1}{N}\langle \tr G(z) \rangle  \, \frac{dz}{d\theta}\, d\theta\right]
= -\frac{1}{\pi}\,\mathrm{Re}\, \left[\frac{i}{2z} iz\right] = \frac{1}{2\pi}
\,.
\label{COErho}
\end{equation}

\end{document}